\documentclass[conference]{IEEEtran}
\usepackage[mathcal]{euscript}
\usepackage{amsmath}
\usepackage{amsthm}
\usepackage{amssymb}
\usepackage{graphicx}
\usepackage{algorithmic}
\usepackage{algorithm}
\usepackage{epsfig}
\usepackage{booktabs}
\usepackage{colortbl}
\usepackage{amsmath}
\usepackage{xcolor}
\usepackage{graphicx}
\usepackage{cite} 

\providecommand{\abs}[1]{\lvert#1\rvert}				
	
\providecommand{\abs}[1]{\lvert#1\rvert}					 

\providecommand{\mbf}[1]{\mathbf{#1}}						 
\providecommand{\wt}[1]{\widetilde{#1}}					 
\providecommand{\mbfwt}[1]{\wt{\mathbf{#1}}}				 
\providecommand{\mc}[1]{\mathcal{#1}}				 
\providecommand{\bsym}[1]{\boldsymbol{#1}}				 
\providecommand{\bsymwt}[1]{\widetilde{\boldsymbol{#1}}}	 

\DeclareMathOperator*{\argmax}{arg \, max}

\IEEEoverridecommandlockouts

\begin{document}

\title{Channel Modeling between Seaborne MIMO Radar and MIMO Cellular System}

\author{\begin{tabular}{c}
 Awais Khawar, Ahmed Abdelhadi, and T. Charles Clancy \\
\{awais, aabdelhadi, tcc\}@vt.edu \\
Ted and Karyn Hume Center for National Security and Technology\\
Bradley Department of Electrical and Computer Engineering\\
Virginia Tech, Arlington, VA, 22203, USA
\end{tabular}
\thanks{This work was supported by DARPA under the SSPARC program. Contract Award Number: HR0011-14-C-0027. The views, opinions, and/or findings contained in this article are those of the authors and should not be interpreted as representing the official views or policies of the Department of Defense or the U.S. Government.

Approved for Public Release, Distribution Unlimited
}

}

\maketitle

\begin{abstract}
Spectrum sharing between radars and cellular systems is an emerging area of research. In this paper, we model channel between a seaborne MIMO radar and MIMO cellular system. We model a 2D channel to capture the azimuth aspect of the spectrum sharing scenario. Our channel modeling methodology allows MIMO radar to place accurate nulls in the azimuth location of base stations (BS), thus, protecting them from harmful radar interference. We use a projection based approach, where radar waveform is projected onto null space of channel, for mitigating radar interference to BSs. This is also known as an approach based on eigen-nulling which is different from spatial-nulling commonly employed by radars. We show through simulations that the proposed spatial channel model allows eigen-nulling which performs superior to traditional spatial-nulling for interference mitigation. The proposed channel model can be leveraged to use eigen-nulling that enhances target detection and beampattern resolution of MIMO radar while mitigating interference to BSs.

\end{abstract}


\section{Introduction}

Recent years have seen tremendous growth in wireless data traffic due to the ubiquitous use of smart phones and tablets. Cellular network operators are finding it hard to cope up with the growing user data requirements given the limited amount of spectrum available to them. Up till now, in order to meet bandwidth requirements, regulators usually free up spectrum held by various incumbents by relocating them to some other band. However, not only this process is time consuming it is also very expensive for incumbents to relocate. That is why regulators are now looking into the option of sharing spectrum across government agencies and commercial services. For example, the Federal Communications Commission (FCC) is considering to allow commercial wireless operation in the 3.5 GHz radar band \cite{FCC12_SmallCells}. This initiative can result in interference issues for radars and communication systems. This letter addresses this issue by modeling channel between radar and communication systems and studies the impact of a proposed interference mitigation scheme.

Channel modeling for wireless communications is a fundamental area of research as it allows performance evaluation of transmission techniques. A comprehensive introduction to SISO and MIMO wireless channel modeling, propagation modeling, and statistical description of channels can be found in \cite{Mol10} and references there in. However, the efforts so far have been limited to model channels between wireless communication systems and to the best of our knowledge no work exists on channel modeling between wireless communication and radar systems. In order to study performance evaluation and interference mitigation techniques in a spectrum sharing scenario between radar and communication systems the problem of channel modeling is of prime importance.

In this letter, we formulate a 2D MIMO channel model between MIMO radar and MIMO cellular system. We consider a 1D antenna array at both the radar and cellular system. Using our proposed channel model we demonstrate the efficacy of eigen-nulling \cite{KAC+14DySPANProjection} over spatial-nulling \cite{VT14} of radar interference at cellular systems. The rest of this letter is organized as follows. Section \ref{sec:models} briefly presents MIMO radar architecture and spectrum sharing scenario. Section \ref{sec:channel} models MIMO channel between seaborne radar and cellular system. Section \ref{sec:nulling} discusses the superiority of eigen-nulling over spatial-nulling approaches and provides numerical examples. Section \ref{sec:conc} concludes the letter.

\section{Spectral-Coexistence Models}\label{sec:models}
In this section, we briefly introduce the fundamentals of MIMO radar and discuss our radar-communication system spectrum sharing scenario.

 \subsection{MIMO Radar}
In the last decade, MIMO radar has emerged as a strong contender for the replacement of legacy radar systems which utilize electronic steering and are commonly known as phased array radar systems. MIMO radars have a distinct advantage in terms of waveform diversity over phased array radar systems as former transmit multiple waveforms that can be chosen freely while later transmits scaled versions of a single waveform. Therefore, in this paper we consider a MIMO radar with $M$ antenna elements and denote samples of baseband equivalent transmitted waveform as $\left\{\mbf{x}(n)\right\}^{L}_{n=1}$. We choose orthogonal waveforms, because of their optimality \cite{LS08}, for which the signal coherence matrix 
can be written as 
\begin{equation*}
\mbf{R}=\frac{1}{L} \sum^L_{n=1} \mbf{x}\left(n\right) {{\mbf{x}}^H}\left(n\right) = \mbf I
\end{equation*}
where $n$ is the time index and $L$ is the total number of time samples. The signal received from a single point target at an angle $\theta$ can be written as \cite{KAC14_TDetect}
\begin{equation}
\mbf y(n) = \alpha \, \mbf A(\theta) \,  {\mbf {x}}(n) + \mbf w(n) \label{eqn:rxRadar}
\end{equation}
where $\alpha$ represents the complex path loss including the propagation loss and the coefficient of reflection and $\mbf A (\theta)$ is the transmit-receive steering matrix.
%

\subsection{Spectral Coexistence Scenario}
We consider a scenario in which a cellular system is deployed in the radar band, i.e, it shares the same channel as that of a monostatic ship-borne MIMO radar system. In this letter, we assume a Frequency Division Duplex (FDD) cellular network in which BSs are operating in the radar band and UEs are operating at non-radar frequencies. Therefore, we focus on interference caused by the radar operation to the base stations and not to the users and model channel for this scenario. Without loss of generality, we consider a single cell or a BS with $N$ transmit and receive antennas, then the signal received by the BS on the uplink, in the presence of radar, is given by 
\begin{equation}
{\mbf{r}}=\sum_{i=1}^{\mc K} \mbf G_i \mbf s_i + {\mbf{H}} \mbf x + \mbf{n} 
\end{equation}  
where $\mc K$ is the number of users in the cell, $\mbf G_i$ is the channel gain between the BS and the $i^{\text{th}}$ user, $\mbf H$ is the channel gain between the BS and the radar, $\mbf x$ is the interfering signal from the MIMO radar which we want to mitigate by designing channel model and then projecting radar signal onto null space of the channel \cite{S.SodagariDec.2012, GhorbanzadehMilcom2014, KAC14_TWS, KAC14_TDetect, Channel3D, ChannelLOS, KAC14_QPSK, KAC+14DySPANProjection, KAC14_Milcom, KAC14DySPANWaveform, KAC14ICC, KAC+14ICNC, SAC+15, SKA+14DySPAN}. 
   
\section{Physical Modeling of Channel}\label{sec:channel}

In this letter, we model the physical channel between MIMO radar and MIMO communication system 
and limit ourselves to model a deterministic channel which will also serve as a basis to extend it to a statistical channel in future.

Consider a scenario in which a BS is an area free of reflectors or scatterers and only direct paths exist between radar and BS antenna pairs. This channel consists of only direct line-of-sight paths between the radar transmit and communication system receive antennas. This scenario is typical of BS antennas mounted on building tops or outer walls of buildings in littoral areas to serve users on a beach, for example. At the BS, the inter-element spacing between antennas is $\Delta_N \lambda_c$, where $\Delta_N$ is the normalized BS antenna separation and $ \lambda_c$ is the carrier wavelength. Similarly, at the radar, the inter-element spacing between antennas is $\Delta_M \lambda_c$, where $\Delta_M$ is the normalized radar antenna separation. Moreover, it is assumed that the radar and BS are sufficiently apart, since, we are interested in channel between a seaborne radar and an on-shore cellular BS. Then, the baseband channel gain between the $k^{th}$ radar transmit antenna and the $i^{th}$ BS receive antenna is as follows:
\begin{equation}\label{eqn:hik}
h_{ik}={a}{\exp\left(-j2 \pi  d_{ik}/\lambda_c \right)}
\end{equation}
where $d_{ik}$ is the distance between the radar and the BS antennas, and $a$ is the attenuation along the line-of-sight path which is assumed to be equal for all antenna pairs. Due to the geometry of the model, it is safe to assume that the antenna array sizes are much smaller than the distance between the transmitter and the receiver, then to a first-order we can write
\begin{equation}\label{eqn:dik}
d_{ik}=d+\left(i-1\right)\Delta_N \lambda_c \cos\phi_N-\left(k-1\right)\Delta_M \lambda_c \cos\phi_M 
\end{equation}
where $d$ is the distance between radar transmit antenna 1 and BS receive antenna 1; $\phi_N$ and $\phi_M$ are the angles of incidence of the line-of-sight path on the radar and BS antenna arrays, respectively. The quantities $\left(i-1\right)\Delta_N \lambda_c \cos\phi_N$ and $\left(k-1\right)\Delta_M \lambda_c \cos\phi_M$ are respectively the displacements of receive/transmit antenna $i/k$ from receive/transmit antenna $1$ in the direction of the line-of-sight. For the simplification of notations lets define $\Omega_M \triangleq \cos\phi_M$ and $\Omega_N \triangleq \cos\phi_N$. 
Therefore, the channel matrix can be written as
\begin{equation} \label{dist_eq}
\mathbf{H}=a\sqrt{{N}M}\exp\left(-j2 \pi  \frac{d}{\lambda_c} \right) \mathbf{e}_N\left(\Omega_N\right) \mathbf{e}_M^*\left(\Omega_M\right) 	
\end{equation}
where unit spatial signature $\mathbf{e}_l$, l=\{N,M\}, is defined to be
\begin{eqnarray}
\mathbf{e}_l\left(\Omega_l\right)&=&\frac{1}{\sqrt{l}} 
\begin{bmatrix}
1\\
\exp\left(-j2 \pi \Delta_l \Omega_l \right)\\
\exp\left(-j2 \pi 2 \Delta_l \Omega_l \right)\\
\vdots \\
\exp\left(-j2 \pi \left(l-1\right) \Delta_l \Omega_l \right)
\end{bmatrix}\cdot
\end{eqnarray}
   
Since we are considering to model propagation channel between a seaborne radar and an on-shore BS, it is of interest to consider multiple paths reflected off sea in addition to the LoS path between the radar and the BS. 
Now, the $i^{\text{th}}$ path has attenuation $a_i$ and makes an angle $\phi_{Mi}$ with the radar transmit antenna array and an angle $\phi_{Ni}$ with the receive antenna array of the BS. The overall channel $\textbf{H}$ is given by
\begin{equation}
\mathbf{H}=\sum^L_{i=1} a_i \mathbf{e}_N\left(\Omega_{Ni}\right) \mathbf{e}_M^*\left(\Omega_{Mi}\right)
\end{equation}
where
\begin{equation}
a_i=a_i \sqrt{NM}\exp\left( \frac{-j2 \pi d^i}{\lambda_c} \right) 
\end{equation}
$d^i$ is the distance between radar transmit antenna 1 and BS antenna 1 along path $i$. 

In this letter, we restrict ourselves to scenarios where no more than one non line-of-sight (NLoS) component exist or LoS component is much much stronger than all other NLoS components except one which might be bounced of water waves.

\section{Spatial vs. Eigen Nulling}\label{sec:nulling}

In radar signal processing, it is often desirable to steer nulls of the transmit beampattern for clutter and interference mitigation purposes. For this reason beamforming weights are optimally designed to maximize signal-to-interference-plus-noise ratio (SINR) so that target signal can be recovered in the presence of clutter, jammer, and interference sources. The SINR maximization can be achieved for various constraints including the constraint that the target response is unity which is mathematically expressed as
\begin{equation}
\begin{aligned}
& \underset{\mbf{w}}{\text{min}}
& & \mbf w^H \mbf R \mbf w \\
& \text{subject to}
& & \mbf{w}^H \mbf x = 1
\end{aligned}
\end{equation}
where $\mbf w$ is the weight vector and the solution of this optimization problem is known as minimum variance distortionless response (MVDR) beamformer \cite{LS08}
\begin{equation}
\mbf w = \frac{\mbf R^{-1} \mbf x}{\mbf x^H \mbf R^{-1} \mbf x}\cdot
 \end{equation} 
A limitation of this approach is that covariance matrix $\mbf R$ needs to be estimated. The spatial locations of nulls - i.e., location of jammers, clutter, and interference sources - are not known \textit{a priori} in most cases and are obtained only after analyzing the received data. In traditional phased array radars, spatial-nulling locations are determined by first illuminating the desired area of interest and then configuring the beampattern with the appropriate nulls whereas in MIMO radar the transmit beampattern is configured with the help of data-dependent processing schemes. Moreover, estimation of correlation matrix of clutter/interference and multipath has to be taken into account for effective spatial transmit nulling.

In the spectrum sharing scenario under consideration where a ship-borne MIMO radar is sharing spectrum with an onshore MIMO communication systems the environmental geometry can be exploited for effective nulling. If traditional spatial nulling techniques are utilized by MIMO radars, for nulling out interference to communication systems, then the covariance matrix of interference and location of multipath scatterers needs to be estimated. Since this process may lead to estimation errors, the effect of nulling can be undermined thus leading to harmful interference for communication systems \cite{VT14}. On the other hand, if eigen-nulling is utilized, i.e., nulls are placed in location of low eigen modes of the interference channel, then the transmit nulling can be very effective in mitigating interference to BSs. Assuming, the MIMO radar has channel state information, through feedback from BSs, we can perform singular value decomposition (SVD) of interference channel $\mbf H$ as 
\begin{equation}
\mbf H = \mbf U \bsym \Sigma \mbf V^H.
\end{equation}
Defining
\begin{equation}
\bsymwt \Sigma \triangleq \text{diag} (\wt \sigma_{1}, \wt \sigma_{2}, \ldots, \wt \sigma_{p})
\end{equation}
where $p \triangleq \min (N,M)$ and 
$\wt \sigma_{1} > \wt \sigma_{2} > \cdots > \wt \sigma_{q} > \wt \sigma_{q+1} = \wt \sigma_{q+2} = \cdots = \wt \sigma_{p} = 0$ are the singular values of $\mbf H$. Next, we define
\begin{equation}
\bsymwt {\Sigma}^\prime \triangleq \text{diag} (\wt \sigma_{1}^\prime,\wt \sigma_{2}^\prime, \ldots, \wt \sigma_{M}^\prime)
\end{equation}
where
\begin{align}
\wt \sigma_{u}^\prime \triangleq
\begin{cases}
0, \quad \text{for} \; u \leq q,\\
1, \quad \text{for} \; u > q.
\end{cases}
\end{align}
The above definitions lead us to our projection matrix defined as
\begin{equation}\label{eqn:ProjDefinition}
\mbf P \triangleq \mbf V \bsymwt \Sigma^\prime \mbf V^H.
\end{equation}
Then, in order to obtain the signal which nulls interference to BSs we project radar signal onto null space of BS channel $\mbf H$ via
\begin{equation}
\mbfwt x = \mbf P \mbf x
\end{equation}
where $\mbfwt x$ is the null space projected radar waveform that is not going to cause interference to BS as the radar waveform is now in the low eigen modes of the interference channel $\mbf H$. 

In the following section we provide two examples which demonstrate the efficacy of eigen-nulling over spatial-nulling schemes.

\subsection{Beampattern}

In this section, we compare transmit-receive beampatterns of MIMO radar waveform subject to eigen- and spatial-nulling. The transmit and receive beampatterns are a means to measure the beamformer's response to a target located at an angle $\theta$, when the beam is steered digitally to a direction $\theta_D$. 
The composite transmit-receive pattern can be expressed as \cite{LS08}
\begin{equation} \label{bptxrx}
G (\theta,\theta_D)=K \dfrac{\abs{\mbf a^H(\theta) \mbf R^T \mbf a(\theta_D)}^2}{\mbf a^H(\theta_D) \mbf R^T \mbf a(\theta_D)}  \dfrac{\abs{\mbf a^H(\theta) \mbf a(\theta_D)}^2}{M}
\end{equation}
where $K$ is the normalization constant. We restrict ourselves to study composite transmit-receive beampatterns because of their advantages including narrower null-to-null beamwidths and smaller sidelobes. Moreover, composite beampatterns are unambiguous due to the presence of both transmit and receive beampatterns. An ambiguity in one beampattern, for example grating lobes, is resolved by the other \cite{LS08}.

\vspace{3mm}

\begin{figure}
\centering
	\includegraphics[width=\linewidth]{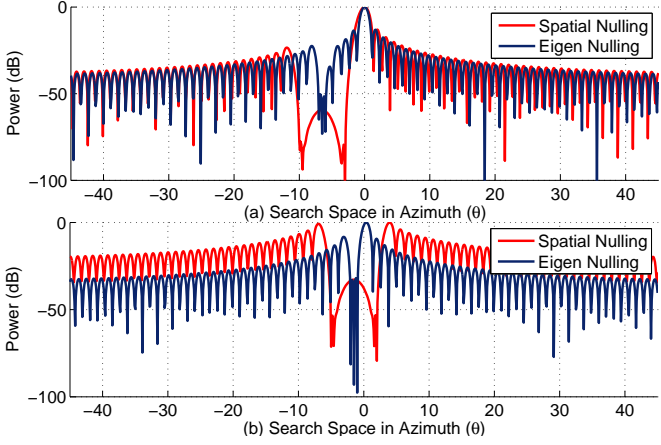} 
	\caption{Analysis of beampattern for spatial- and eigen-nulling approaches. Eigen-nulling results in accurate nulling and target localization for targets that are both far and near to BS locations as compared to spatial-nulling.}
		\label{fig:BP}
\end{figure}

\noindent
\textbf{Numerical Example:} In Figure \ref{fig:BP}, we compare transmit nulling approaches based on spatial-nulling and eigen-nulling. We are interested in understanding the effects of eigen-nulling and spatial-nulling for interference mitigation on the beampattern of the spectrum sharing MIMO radar. In Figure \ref{fig:BP}a, we consider the case when target and BSs are far away in spatial domain, i.e, we assume the target is at an angle $\theta = 0^\circ$, BS has a direct LoS angle at $\theta=-7^\circ$ and a NLoS component is at $\theta = -6^\circ$, this is based on the CSI exchanged between radar and BS. In the absence of such an information, traditional spatial-nulling techniques are employed that have to estimate spatial locations that need to be blocked and thus in many cases a wider range of angles have to be nulled to protect BS from radar interference. Hence, for this example we block spatial angles from $\theta = -10$ to $-3$. It can be noticed that using our proposed 2D channel model we can leverage eigen-nulling which results in accurate nulling of interference at BS location and target localization. Although, spatial-nulling also results in accurate localization of target but has a large nulled area in its search space. In Figure \ref{fig:BP}b, we consider the 
case when target and BS are very close in spatial domain, i.e, we assume the target is at an angle $\theta = 0^\circ$, BS has a direct LoS angle at $\theta=-3^\circ$ and a NLoS component is at $\theta = -2^\circ$, this is based on the CSI exchanged between radar and BS. If traditional spatial-nulling techniques are employed then a wider range of angles have to be nulled to protect BS from radar interference. Hence, for this example we block spatial angles from $\theta = -5$ to $-2$. It can be noticed that using our proposed 2D channel model we can leverage eigen-nulling which results in accurate nulling of interference at BS location and target localization for even closely located BSs and target. On the other hand, spatial-nulling fails to accomplish target localization due to the nulls placed in the direction of target. Hence, 
using our proposed 2D channel model we can leverage eigen-nulling which results in accurate nulling of interference and target localization for both close and far away targets.

\subsection{Probability of Detection}
In this section, we compare the target detection performance of radars that utilize spatial nulling vs. eigen nulling approaches for interference mitigation. The generalized likelihood ratio test (GLRT) for spectrum sharing MIMO radars is given by \cite{KAC14_TDetect}
\begin{equation}
\label{eq:ML}
L_{\mbf y}(\hat{\theta}_{\text{ML}}) = \argmax\limits_{\theta} \frac{\left| \mathbf{a}^H(\hat{\theta}_{\text{ML}}) \mbf E \mathbf{a}^*(\hat{\theta}_{\text{ML}})\right|^2}{M \mathbf{a}^H(\hat{\theta}_{\text{ML}}) \mathbf{R}^T \mathbf{a}(\hat{\theta}_{\text{ML}})} \mathop{\gtrless}^{\mc H_1}_{\mc H_0} \delta
\end{equation}
where the asymptotic statistics of $L(\hat{\theta}_{\text{ML}})$ is given by
\begin{align}
L(\hat{\theta}_{\text{ML}}) \sim 
\begin{cases}
\mc H_{1}: \chi^2_2\left(\dfrac{|\alpha|^2}{\sigma^2_n} | \mathbf{a}^H({\theta}) \mathbf{R}^T \mathbf{a}({\theta})|^2\right) \\
\mc H_{0}:\chi^2_2
\end{cases}
\end{align}
where $\hat{\theta}_{\text{ML}}$ is the maximum likelihood (ML) estimate of $\theta$, $\mc H_{0}$ is the null hypothesis, $\mc H_{1}$ is the alternate hypothesis, $\delta$ is the detector threshold, $\chi^2_2(\cdot)$ is the noncentral chi-squared distributions with two degrees of freedom, and $\chi^2_2$ is the central chi-squared distributions with two degrees of freedom.
The probability of detection is given by
\begin{align}
P_{\text{d}} &= P(L(\mbf y) > \delta | \mc H_1) \\
P_{\text{d}} &=1 - \mathcal{F}_{\chi^2_2(\rho)} \left( \mathcal{F}^{-1}_{\chi^2_2} (1-P_{\text{FA}}) \right)
\end{align}
where $\mathcal{F}_{\chi^2_2(\rho)}$ is the noncentral chi-squared distribution function with two degrees of freedom and noncentrality parameter $\rho$ and $P_{\text{FA}}$ is the probability of false alarm. 
\vspace{3mm}

\begin{figure}
\centering
	\includegraphics[width=\linewidth]{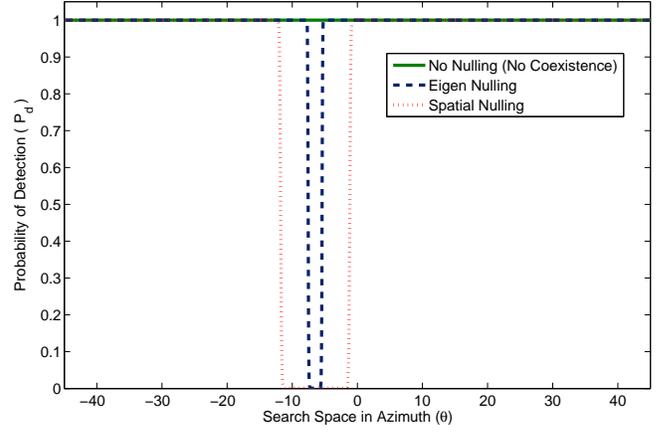} 
	\caption{Comparison of probability of detection for spatial- and eigen-nulling approaches. Eigen-nulling results in minimum loss in search space as compared to spatial-nulling.}
		\label{fig:Pd}
\end{figure}

\noindent
\textbf{Numerical Example:} In Figure \ref{fig:Pd}, we study the target detection capability of the MIMO radar in a spectrum sharing environment. We are interested in understanding the effects of eigen-nulling and spatial-nulling for interference mitigation on the radar's targets detection probability. We assume the target is at an angle $\theta = 0^\circ$, BS has a direct LoS angle at $\theta=-7^\circ$ and a NLOS component is at $\theta = -6^\circ$, this is based on the CSI exchanged between radar and BS. For spatial-nulling we block angles from $\theta = -10$ to $-3$. From Figure \ref{fig:Pd} it can be noticed that when eigen-nulling is performed we have zero detection probability for only the angles which strongly interfere with the BS, however, when spatial-nulling is used extra angles are nulled which leads to zero detection probability for even those angles which do not interfere with BS. Thus, our proposed 2D channel model can leverage eigen-nulling which performs much better than spatial-nulling as it provides minimum loss in radar search space as compared to spatial-nulling which blocks significant volume (azimuth space) which is not desirable for radar operations. 

\section{Conclusion}\label{sec:conc}
In this letter we modeled a MIMO channel for a seaborne radar sharing spectrum with an onshore communication system. Using the proposed channel model we showed that eigen-nulling based approaches yield better target localization, accurate interference-nulling, and minimum loss in radar search/detection space as compared to traditional spatial-nulling approaches widely used in radar signal processing.

\bibliographystyle{ieeetr}
\bibliography{IEEEabrv,channel2D}
\end{document}